\documentclass[10pt,twocolumn,preprintnumbers,amsmath,amssymb,nofootinbib
,superscriptaddress]{revtex4}
\usepackage{graphicx,longtable,mathrsfs,color,array}
\usepackage[usenames,dvipsnames]{xcolor} 
\usepackage{amssymb,amsmath,mathtools,mathrsfs,slashed} 
\usepackage{epsfig,subfigure,placeins,float} 
\usepackage{booktabs,longtable,ctable,multirow} 
\usepackage{exscale,relsize} 
\usepackage[normalem]{ulem} 
\usepackage{enumerate}
\usepackage{times, mathptmx} 
\usepackage[colorlinks,linkcolor=blue,citecolor=blue,urlcolor=blue ]{hyperref}

\usepackage[stable]{footmisc}
\usepackage{color}

\newcommand{\jcap}{JCAP}
\newcommand{\mnras}{MNRAS}

\newcommand{\procspie}{Proceedings of the SPIE}

\newcommand{\fevo}{f_{\rm evo}}

\begin{document}

\title{The impact of relativistic effects on cosmological parameter estimation}
\author{Christiane  S. Lorenz}
\email{christiane.lorenz@physics.ox.ac.uk}
\affiliation{Astrophysics, University of Oxford, DWB, Keble Road, Oxford OX1 3RH, UK}
\author{David Alonso}
\affiliation{Astrophysics, University of Oxford, DWB, Keble Road, Oxford OX1 3RH, UK}
\author{Pedro G. Ferreira}
\affiliation{Astrophysics, University of Oxford, DWB, Keble Road, Oxford OX1 3RH, UK}

\date{Received \today; published -- 00, 0000}

\begin{abstract}
Future surveys will access large volumes of space and hence very long wavelength fluctuations of the matter density and gravitational field. It has been argued that the set of secondary effects that affect the galaxy distribution, relativistic in nature, will bring new, complementary cosmological constraints. We study this claim in detail by focusing on a subset of wide-area future surveys: Stage-4 cosmic microwave background experiments and photometric redshift surveys. In particular, we look at the magnification lensing contribution to galaxy clustering and general relativistic corrections to all observables. We quantify the amount of information encoded in these effects in terms of the tightening of the final cosmological constraints as well as the  potential bias in inferred parameters associated with neglecting them. We do so for a wide range of cosmological parameters, covering neutrino masses, standard dark-energy parametrizations and scalar-tensor gravity theories. Our results show that, while the effect of lensing magnification to number counts does not contain a significant amount of information when galaxy clustering is combined with cosmic shear measurements, this contribution does play a significant role in biasing estimates on a host of parameter families if unaccounted for. Since the amplitude of the magnification term is controlled by the slope of the source number counts with apparent magnitude, $s(z)$, we also estimate the accuracy to which this quantity must be known to avoid systematic parameter biases, finding that future surveys will need to determine $s(z)$ to the $\sim$5-10\% level. On the contrary, large-scale general-relativistic corrections are irrelevant both in terms of information content and parameter bias for most cosmological parameters, but significant for the level of primordial non-Gaussianity.
\end{abstract}


\maketitle
\section{Introduction}

In the next decade, we expect to map out the large scale structure of the Universe with exquisite precision. In doing so it will be possible, for the first time, to access information on the largest possible scales -- the scale of the cosmological horizon. It has been shown that, on those scales, a number of general relativistic effects come into play \cite{Bonvin:2005ps,Bonvin:2011bg,Yoo:2010ni,Challinor:2011bk}. Such effects might, conceivably, lead to additional and complementary information to that obtained on the usual scales probed by current surveys ($\lesssim 100\,h^{-1}$ Mpc).

General-relativistic effects are more significant on large scales; unfortunately there are fewer modes to sample and cosmic variance severely limits our ability to detect these effects in the standard way. Indeed, it has been shown that from auto-correlations alone (i.e. from the power spectra of individual tracers) it is impossible to detect these effects with any statistical significance \cite{Alonso:2015uua}, and the only way to measure them is via cross-correlations of data sets, through what has been dubbed the multi-tracer technique \cite{Seljak:2008xr}. It has been shown that a judicious choice of future surveys can be combined to obtain a moderate to high significance detection of general relativistic effects \cite{Alonso:2015sfa,Fonseca:2015laa}.

Common sense would dictate that the various, novel, effects that have been identified need to be taken into account if we are to constrain cosmological parameters from future surveys. Indeed, it has been shown that some of these effects can play a significant role and bias the outcomes of cosmological parameter estimation. We highlight two cases: constraints on primordial non-Gaussianity and the impact of lensing magnification on galaxy number counts.

If primordial fluctuations were non-Gaussian, it has been shown that one should expect corrections in the small $k$ (large wavelength) part of the galaxy power-spectrum through scale-dependent biasing \cite{Dalal:2007cu,Matarrese:2008nc}. This effect, in which the bias parameter gets a correction $\Delta b\propto 1/k^2$, can be confused with some of the general relativistic effects \cite{2012PhRvD..85d1301B}. Thus a correct accounting of both scale dependent biasing and general relativistic effects must be adopted in any analysis of long wavelength modes. 

Alternatively, it has been well established that lensing will affect measurements of the galaxy distribution through, for example, magnification bias \cite{1980ApJ...242L.135T}. Lensing may have a significant effect on all scales and it has been shown that if it is not correctly included, it may lead to significant biases in estimates of cosmological parameters such as the neutrino mass scale \cite{Cardona:2016qxn} or the dark energy equation of state \cite{2014MNRAS.437.2471D}.

In this paper we will systematically explore the role that general-relativistic effects (and large scale modes) play on cosmological parameter constraints. Our focus will be on the importance of lensing correction  (following up on the work of \cite{2014MNRAS.437.2471D,Cardona:2016qxn}) and on the combined general relativistic corrections to galaxy number counts. We will use a Fisher matrix analysis to quantify the importance of these effects on the forecast errors and on the potential measurement bias of cosmological parameters from a selection of Stage IV experiments. We will be comprehensive in our analysis of cosmological parameters in that we will include the standard set of $\Lambda$CDM parameters but also encompass a time-varying equation of state for dark energy, the mass of neutrinos, primordial non-Gaussianity and scalar-tensor extensions to the theory of gravity.

We structure this paper as follows. In Section \ref{sec:observables} we briefly recap the effects that we will be studying and discuss the methodology that we will use. In Section \ref{sec:methodology} we explain the various parts that go into the Fisher matrix formalism for forecasting and how it can be used to quantify potential biases in the analysis. We then, in Section \ref{sec:results}, systematically work through the different combination of data sets and cosmological parameters to build up a comprehensive analysis of the role these effects will play in future surveys. In Section \ref{sec:discussion} we discuss the results of our analysis.


\section{Observables and Large Scale Effects}
\label{sec:observables}
The goal of modern cosmology is to map out the large-scale structure of the Universe. To do so, observers try to quantify the statistical properties of the distribution of matter by either studying the spatial distribution of bright objects (such as galaxies) or diffuse gas, or by measuring the effect of gravitational potentials on the propagation of light emitted by distant sources. Key to such observations is to accurately characterize the redshifts and directions of photons that propagate from cosmological distances to observing instruments. From these properties, one can infer the density perturbations, observable volume distortions and perturbed photon paths.

A key quantity is the fluctuation in the number density of galaxies at a particular solid angle and at a particular redshift. The corresponding observable, $\Delta_N(z,{\bf \widehat{n}})$, consists of a number of terms which can be schematically written as \cite{Challinor:2011bk,Bonvin:2011bg}
\begin{eqnarray}\label{eq:terms1}
\Delta_N\equiv\Delta^D+\Delta^{RSD}+\Delta^{L}+\Delta^{GR}
\end{eqnarray}
where ``$D$'' stands for density perturbations, ``$RSD$'' stands for redshift space distortions, ``$L$'' stands for lensing magnification and ``$GR$'' stands for general-relativistic corrections. The first three terms are dominant and play a role on all scales -- they are, at most, weighted by linear factors of ${\cal H}/k$ where ${\cal H}$ is the conformal Hubble factor and $k$ is the wavenumber of the perturbation. The general-relativistic corrections include large-scale velocity terms and terms involving the gravitational potentials and their derivatives (akin to the integrated Sachs-Wolfe \cite{1967ApJ...147...73S} effect and the Shapiro time delay \cite{1964PhRvL..13..789S}, found in other settings). The ``$GR$'' terms are typically weighted by factors of 
$({\cal H}/k)^2$. The exact expressions for all these terms can be found in Appendix \ref{app:GRterms}.

Redshift space distortions, or the ``Kaiser effect'' (see~\cite{kaiser1987})  are currently the method {\it par excellence} for measuring the growth rate of structure, $f=d\ln\delta_M/d\ln a$ (where $\delta_M$ is the matter density contrast and $a$ is the scale factor) \cite{2017MNRAS.466.2242B}. These distortions arise from the peculiar velocity sourced by the local gravitational potential which induce shifts in the relationship between the distance and redshift of any particular galaxy. The interplay between the RSD term and the density contrast involves the clustering bias, $b$, which relates the number density with fluctuations in the comoving-gauge matter perturbations. As such, measuring the growth rate will involve assumptions about the tracer being considered and can, potentially, be amenable to multi-tracer techniques \cite{Seljak:2008xr}.

We will pay particular attention to the magnification term, the most significant effect after RSDs and already well measured by multiple analyses \cite{1979ApJ...227...30S,2003ApJ...589...82G,2005ApJ...633..589S,2012MNRAS.426.2489M,2016MNRAS.457.3050C,2016arXiv161110326G}. This magnification bias depends on the slope of the physical number density of sources, ${\bar {\cal N}}(\eta,L>L_*)$, as a function of conformal time $\eta$ and intrinsic luminosity $L_*$, as:
\begin{eqnarray}
\label{eq:sz}
s\equiv\frac{5}{2}\frac{\partial \ln {\bar {\cal N}}}{\partial \ln L_*}.
\end{eqnarray}
This correction arises because of the presence of matter overdensities along the photon path, on the one hand stretching the observed separation between galaxies (and therefore surpressing the observed number density) and on the other hand boosting the observability of faint galaxies which otherwise would have fallen below the detection threshold \cite{Hui:2007cu}. As we shall see (and as was pointed out in \cite{2014MNRAS.437.2471D,Cardona:2016qxn}), this term can play a significant role in biasing the estimates of cosmological parameters.

The GR effects are subdominant and only really emerge on the largest scales (as can be seen in Appendix \ref{app:GRterms}). There are a few main things to note which will become important when discussing the methodology and results. First of all, the fact that they are weighted by 
$({\cal H}/k)^2$ means that they come in with a similar scale dependence as the scale dependent bias arising from primordial non-Gaussianity. Second, some of the terms depend  on the slope of the background number density of sources as a function of time, the {\it evolution bias}:
\begin{eqnarray}
f_{\rm evo}\equiv \frac{\partial \ln (a^3 {\bar {\cal N}})}{\partial \ln a}
\end{eqnarray}
Given this, it has been shown \cite{Alonso:2015sfa,Fonseca:2015laa} that GR effects are amenable to the use of multitracer techniques for mitigating cosmic variance and that, with the appropriate choice of future data sets, it may be possible to detect them at the $\sim10\sigma$ level.

\section{Methodology}
\label{sec:methodology}

\subsection{The space of parameters}
\label{ssec:methodology.parameters}
In this work, we will consider a number of different cosmological models in order to make a broad and general statement about the impact of the lensing and general-relativistic effects on the estimation of cosmological parameters. 

As a first model we choose the standard extension to $\Lambda$CDM including non-zero neutrino masses $\sum m_\nu$ and a time-varying equation of state for dark energy. The latter is parametrized by $w_0$ and $w_a$ \cite{2001IJMPD..10..213C} as $w(a)=w_0+(1-a)w_a$. This model also includes the standard cosmological parameters (fractional density of dark matter $\Omega_{\rm cdm} h^2$ and baryons $\Omega_{\rm b}h^2$, the local normalized expansion rate $h$, the amplitude of primordial scalar perturbations $A_{\rm s}$, the scalar spectral index $n_{\rm s}$ and the optical depth to reionization $\tau$). For these parameters, apart from $\tau$, we will take the best-fit values from the \textit{Planck} 2015 analysis \cite{planck2015xiii} as our fiducial cosmology. We will also take a fiducial $\tau=0.06$ from the latest measurement from \textit{Planck}~\cite{Aghanim:2016yuo}. So far, only lower and upper limits for $\sum m_\nu$ are known. Whereas the currently best upper limits on $\sum m_\nu$ come from cosmology~\cite{Palanque-Delabrouille:2013gaa,planck2015xiii,Vagnozzi:2017ovm}, the mass differences between the neutrino mass eigenstates have been measured in neutrino oscillation experiments. Here we will conservatively use $\sum m_\nu=0.06$ eV as a fiducial value for the total neutrino mass, corresponding approximately to the current lower bound on the total neutrino mass sum from summing the mass differences~\cite{patrignani2016}. Finally, our fiducial dark energy equation of state will correspond to a cosmological constant with $w_0=-1$ and $w_a=0$.

Our second model will extend the previous one with the dimensionless parameter $f_{\rm NL}$ that describes the amount of non-Gaussianity in the primordial density field produced in many inflation scenarios. Specifically we will focus on the case of local non-Gaussianity \cite{Komatsu:2001rj}, in which $f_{\rm NL}$ is defined through
\begin{equation}
\Phi({\bf x})=\Phi_G({\bf x})+f_{\rm NL}(\Phi^2_G({\bf x})-\langle\Phi^2_G\rangle),
\end{equation}
where $\Phi$ is the primordial gravitational potential and $\Phi_G$ is a Gaussian random field. Thus, the primordial gravitational potential can be described as the sum of a linear term and a non-linear one. 
The current constraint on the local value of $f_{\rm NL}$ from the \textit{Planck} satellite is $2.5\pm 5.7$~\cite{planck2015xvii}. Although measurements of the cosmic microwave background anisotropies will be most helpful in determining the value of $f_{\rm NL}$~\cite{Verde:1999ij}, its effects on large-scale structure \cite{Dalal:2007cu,Matarrese:2008nc} are one of the most promising ways to improve current constraints. 
More specifically, primordial non-Gaussianity induces a correction in the Gaussian bias $b_X^G$ of each tracer $X$~\cite{Dalal:2007cu,Matarrese:2008nc}
\begin{equation}
\Delta b_X(z,k)=3f_\mathrm{NL}\frac{[b_X^G(z)-1]\Omega_mH_0^2\delta_\mathrm{c}}{(T(k)D(z)k^2)}
\end{equation}
where $\Omega_m$ is the fraction of the matter density of the total energy density in the Universe, $\delta_\mathrm{c}\simeq1.686$, $D(z)$ is the linear growth factor, $H_0$ the value of the Hubble constant today and $T(k)$ the matter transfer function. As fiducial value for $f_{\rm NL}$ we choose $f_{\rm NL}=0$. 

In these two models General Relativity is still the underlying theory of gravity. For our third model, and in order to explore the role of relativistic effects in constraining deviations from GR, we will consider scalar-tensor theories within the Horndeski class of models~\cite{horndeski,Deffayet:2009wt}. As proposed by \cite{Bellini:2014fua}, these models can be described through a number of general time-dependent functions $\alpha_M$, $\alpha_K$, $\alpha_B$, $\alpha_T$ and $M_*$ in addition to the standard $\Lambda$CDM parameters (we refer the reader to the reference above for further details). These functions parametrize the time variation of Newton's constant ($M_*$ and $\alpha_M$), the form of the scalar kinetic term $\alpha_K$, the mixing between the scalar field and the scalar perturbations $\alpha_B$ and the speed of propagation of tensor modes $\alpha_T$. In order to curb the freedom allowed by this parametrization we constrained the time-dependence of the $\alpha$ functions to be of the form:
\begin{equation}
  \alpha_X(z)=c_X\,\frac{\Omega_{\rm DE}(z)}{\Omega_{\rm DE}(z=0)},
\end{equation}
where $\Omega_{\rm DE}(z)$ is the fractional energy density of the dark energy component. Furthermore, as in \cite{Bellini:2015xja,Alonso:2016suf} we will only consider $c_M$, $c_B$ and $c_T$ as free parameters, since $c_K$ and $M_*$ cannot be constrained by current \cite{Bellini:2015xja} or future data\footnote{Fortunately these parameters are not significantly degenerate with the rest, and therefore can be safely kept fixed without affecting the forecast constraints \cite{Alonso:2016suf}.}. As fiducial values we chose $c_{\rm{B}}=0.05$, $c_{\rm M}=-0.05$ and $c_{\rm{T}}=-0.05$, in order to stay close to $\Lambda$CDM as a fiducial cosmology while avoiding the singularity at $c_X\equiv0$.

\subsection{Fisher matrix forecasting formalism}\label{ssec:methodology.fisher}
We produce our forecasts using a Fisher matrix approach. We follow the formalism of \cite{Alonso:2015uua}, which incorporates the joint constraining power of multiple experiments and tracers of the matter distribution\footnote{The software used to produce these forecasts can be found at \url{https://github.com/damonge/GoFish}.}. Each tracer contains a set of sky maps corresponding to e.g. different redshift bins or the different Stokes polarization parameters in a CMB experiment. In total, the combination of all tracers will observe a number of $N_{\rm maps}$ maps that can be described by their harmonic coefficients $a_{\ell m}^{a,i}$, where $a$ and $i$ label the tracer and map number respectively. We group these harmonic coefficients into a vector ${\bf a}_{\rm \ell m}$ and define the power spectrum ${\sf C}_\ell$ as the covariance of this vector: 
\begin{equation}
\left\langle{\bf a}_{\ell m}{\bf a}_{\ell'm'}^*\right\rangle=\delta_{\ell\ell'}\delta_{mm'}{\sf C}_{\ell}
\end{equation}
We assume that the $a_{\ell m}^{a,i}$ are Gaussian-distributed and that thus their likelihood is given by
\begin{equation}
\label{eq:like}
-2\ln\mathcal{L}=\sum_\ell f_{\rm sky}\frac{2\ell+1}{2}\left[\sum_{m=-\ell}^\ell\frac{{\bf a}_{\ell m}^\dag{\sf C}_\ell^{-1}{\bf a}_{\ell m}}{2\ell+1}+\ln(\mathrm{det}[2\pi{\sf C}_\ell])\right]
\end{equation}
By expanding this likelihood around the maximum we find that the covariance of the maximum-likelihood estimate of a set of parameters $\theta_\alpha$ can be approximated by the inverse of the Fisher matrix ${\sf F}_{\alpha\beta}$. This matrix can be computed as:
\begin{equation}\label{eq:fisher}
{\sf F}_{\alpha\beta}=\sum_{l=2}^{l_{\rm max}}f_{\rm sky}\frac{2\ell+1}{2}
\,{\rm Tr}\left[(\partial_\alpha {\sf C}_\ell) {\sf C}_\ell^{-1}(\partial_\beta{\sf C}_\ell) {\sf C}_\ell^{-1}\right]
\end{equation}
where $f_{\rm sky}$ is the fraction of the sky observed.

The power spectra were computed with a modified version of {\sf{CLASS}}~\cite{DiDio:2013bqa,2011arXiv1104.2932L,2017JCAP...08..019Z}, and the derivatives in Eq. \ref{eq:fisher} were estimated via central finite differences: 
\begin{equation}
\partial_\alpha f=\frac{f(\theta_\alpha+\delta\theta_\alpha)-f(\theta_\alpha-\delta\theta_\alpha)}{2\delta\theta_\alpha}+{\mathcal O}(\delta\theta_\alpha^3).
\end{equation}
The final parameter uncertainties are computed from the inverse of ${\sf F}$.

Besides the parameter uncertainties for a given setup, we also estimate the bias on those parameters arising from neglecting to account for a given relativistic effect in the theoretical calculation of the power spectra. In order to do so, we follow a similar method based on expanding the likelihood around the maximum. The approach is similar to that of \cite{Huterer:2004tr,2006MNRAS.366..101H,Amara:2007as}. As in \cite{Cardona:2016qxn}, we compute an ``observed'' power spectrum ${\sf C}_\ell^{\rm obs}$, where all relevant effects are included in the calculation, and a ``theoretical'' power spectrum ${\sf C}_\ell^{\rm th}$, where a given effect (e.g. lensing magnification or the contribution of large-scale GR effects) is not incorporated. Likewise, we define $\theta_{\rm inf,\alpha}$ as the ``inferred'' values of the cosmological parameters from the incorrect likelihood, and $\theta_{\rm true,\alpha}$ as the true underlying parameters. The maximum-likelihood value for $\theta_\alpha$ is derived by maximising the likelihood in Eq. \ref{eq:like}, and therefore we obtain:
\begin{equation}
\begin{split}
\langle\partial_\alpha\chi^2(\theta_{\rm inf})\rangle\approx & \langle\partial_\alpha\chi^2(\theta_{\rm true})\rangle \\
& +\langle\partial_\alpha\partial_\beta\chi^2(\theta_{\rm true})\rangle(\theta_{\rm true}-\theta_{\rm inf})=0
\end{split}
\end{equation}
Taking $v_\alpha=-\langle\partial_\alpha\chi^2(\theta_{\rm inf})\rangle$ and approximating\footnote{see appendix~\ref{app:Fisher bias calculation} for details}
\begin{equation}
\langle\partial_\alpha\partial_\beta\chi^2(\theta_{\rm inf})\rangle\approx{\sf F}_{\alpha\beta} \,,
\end{equation}
we obtain the bias on each cosmological parameter $\theta_\alpha$:
\begin{equation}
\label{eq:bias}
\Delta\theta_\alpha=({\sf F}^{-1}\cdot {\bf v})_\alpha,
\end{equation}
where the entries of ${\bf v}$ are given by
\begin{equation}
\label{eq:v}
v_\alpha=\sum_{\ell=2}^{\ell_{\rm max}} f_{\rm sky}\frac{2\ell+1}{2}{\rm Tr}\left[(\partial_\alpha{\sf C}_\ell){\sf C}_\ell^{-1}\Delta{\sf C}_\ell{\sf C}_\ell^{-1}\right] \,,
\end{equation}
and $\Delta {\sf C}_\ell={\sf C}_\ell^{\rm obs}-{\sf C}_\ell^{\rm th}$.

\subsection{Upcoming surveys}\label{ssec:methodology.surveys}
We will perform our forecasts for two complementary Stage-IV experiments with optimal area  overlap: CMB S4 and LSST. Together, they will offer at least four different cosmological tracers: CMB primary and lensing, cosmic shear and galaxy clustering, the latter two encompassing several redshift bins. The assumptions used to model these experiments are described here. In all cases we correctly account for all correlations between different tracers.

\subsubsection{CMB Stage 4}\label{sssec:methodology.surveys.s4}
In the mid 2020s, the current ground-based CMB facilities such as Advanced ACTPol \cite{2014JCAP...08..010C}, SPT-3G \cite{2014SPIE.9153E..1PB}, BICEP2/Keck~\cite{Array:2015xqh} or the Simons Array \cite{2016JLTP..184..805S} will be superseded by a CMB Stage 4 (S4) experiment~\cite{abazajian2016}, combining the efforts of multiple ground-based instruments. S4 will be able to derive cosmological constraints from a number of probes, including the primary CMB anisotropies in temperature and polarization, the CMB lensing convergence, Sunyaev-Zel'dovich cluster number counts and other secondary anisotropies. Of these, our forecasts will include the first two, given their relative robustness to astrophysical systematics. Following \cite{calabrese2016} we model S4 as an experiment mapping 40\% of the sky with an rms noise sensitivity of $1\mu{\rm K}$-${\rm arcmin}$ in temperature and a $3\,{\rm arcmin}$ full width at half maximum beam. Given the important systematic uncertainties on large scales faced by ground-based experiments (associated for instance to atmospheric noise or ground pickup), we further assume that S4 will only be able to effectively cover the multipole range $30<\ell<3000$ in temperature and $30<\ell<5000$ in polarization (with the lower small-scale cut in temperature motivated by the effect of astrophysical foregrounds). On $\ell<30$ we supplement S4 with large-scale data from \textit{Planck}~\cite{planck2013} with the corresponding noise level. Although we model the noise contribution to the CMB power spectrum as white, the atmosphere generates a non-trivial noise structure on large scales, especially in temperature. The cosmological parameters considered here are however mostly constrained from the high-$\ell$ CMB power spectrum, and therefore our forecasts should not be strongly affected by this.

It is worth noting that the validity of the Fisher matrix approach can be particularly sensitive to the degeneracies between different parameters (both in terms of predicted uncertainties and biases). Of particular interest are the existing degeneracies between $A_s$, $\tau$, $\Omega_M$ and $\sum m_\nu$, one of the main obstacles to measuring neutrino masses given the currently large uncertainties on $\tau$ from \textit{Planck} \cite{Aghanim:2016yuo,calabrese2016}. In order to verify that our results are not significantly affected by numerical instabilities associated to these degeneracies, we have recalculated our forecasts supplementing S4 on $\ell<30$ with an optimal future satellite mission with a sensitivity of $4\mu{\rm K}$-${\rm arcmin}$ in temperature. This setup is able to reach a cosmic-variance-limited error on $\tau$, and therefore significantly reduce these parameter correlations. Doing this we verified that the results shown in Section \ref{sec:results} are stable with respect to parameter degeneracies.

\subsubsection{Large Synoptic Survey Telescope}\label{ssec:methodology.surveys.lsst}
The Large Synoptic Survey Telescope \cite{2009arXiv0912.0201L} will carry out a 10-year deep and wide imaging survey of the southern sky, reaching a limiting magnitude of $r\sim27$ over $\sim20,000\,{\rm deg}^2$. The use of photometric redshifts (photo-$z$) to obtain approximate radial coordinates will allow LSST to obtain cosmological constraints from a number of probes. These will include tomographic galaxy clustering and cosmic shear, galaxy cluster counts, type Ia supernovae and strong lensing. In particular the complementarity between clustering and lensing make the joint analysis of these two probes the most promising source of cosmological information for LSST, and therefore our forecasts are based on these. We base our modelling of both tracers on the treatment of \cite{Alonso:2015uua}, which we describe briefly below.

\paragraph{Galaxy clustering.} In this case the most relevant observable is the shape of the angular power spectrum or correlation function of the galaxy distribution. The standard way to analyze it will be in terms of tomographic redshift bins, including all auto- and cross-correlations between them. We further separate the clustering sample into two disjoint populations of ``red'' (early-type, ellipticals, high-bias) and ``blue'' (late-type, disks, low-bias) galaxies. The specific models used for the signal and noise power spectra, redshift distributions and nuisance parameters are described in detail in \cite{Alonso:2015uua}.

The relation between the galaxy and matter power spectra is expected to be well-approximated by a linear ``clustering bias'', scale-independent, factor $b(z)$ on large scales. Our forecasts therefore marginalize over the value of this quantity defined, for each galaxy sample, at a discrete set of nodes in redshift (with the full $b(z)$ function reconstructed by interpolating between these nodes, see \cite{Alonso:2015uua} for details). This approximation is, however, bound to fail on small scales, where non-linear, scale-dependent corrections, as well as stochastic contributions, should be taken into account. This makes the analysis of galaxy clustering on small scales very unreliable and often unusable for cosmology. In order to avoid these complications we define, for each redshift bin, angular scale cuts within which the corresponding map is used. At the median redshift of the $i$-th redshift bin $z_i$ we compute a threshold comoving scale $k^i_{\rm max}$ defined as the cutoff scale for which the variance of the linear matter density contrast on larger scales is below a given threshold $\sigma_{\rm thr}^2$, i.e:
\begin{equation}\label{eq:sigma_thr}
  \sigma_{\rm thr}^2=\frac{1}{2\pi^2}\int_0^{k^i_{\rm max}}dk\,k^2\,P(k,z_i).
\end{equation}
This comoving scale is then translated into an angular multipole $\ell^i_{\rm max}=\chi(z_i)\,k^i_{\rm max}$. For our fiducial forecasts we used a threshold variance of $\sigma_{\rm thr}=0.75$.

\paragraph{Cosmic shear.} The effect of weak gravitational lensing observed through the projected shapes of galaxies is a direct, unbiased probe of the intervening matter distribution. As such, cosmic shear observations are a potentially strong cosmological probe. The constraining power of this probe is contained in the power spectrum of the traceless part of the cosmic shear tensor for galaxies lying in a set of photo-$z$ bins. As described in \cite{Alonso:2017dgh}, we model the galaxy sample used for cosmic shear after the so-called ``gold sample'' \cite{2009arXiv0912.0201L}, corresponding to galaxies with magnitude $i<25.3$. We refer the reader to \cite{Alonso:2017dgh} for further details on this sample definition as well as the form of the lensing power spectrum assumed in this analysis. We use a constant minimum scale $\ell_{\rm max}=2000$ for cosmic shear in our forecasts.

Both galaxy clustering and cosmic shear suffer from a number of sources of systematic uncertainties beyond those described above, such as photo-$z$ uncertainties, the effect of intrinsic alignments or baryonic uncertainties in the matter power spectrum. In order to simplify the analysis we have neglected these systematics\footnote{The conservative scale cuts used here have been shown in \cite{Alonso:2016suf} to be robust against the impact of baryonic uncertainties.}. The final constraints on cosmological parameters depend critically on these uncertainties, as well as on the range of angular scales included in the analysis. The absolute forecast constraints on cosmological parameters reported in the next section should therefore not be taken at face value, but rather interpreted in terms of the relative information gain associated to the magnification and relativistic effects, as well as the associated relative biases.


\section{Results}
\label{sec:results}

This section explores the relevance of the magnification bias and the other sub-dominant relativistic corrections to the number counts power spectrum. Here ``relevance'' will be evaluated in terms of both the information content (i.e. constraining power on particular cosmological parameters) and the associated systematic (i.e. possible bias on the same parameters) of these effects. The results will be presented for three different families of parameter spaces. These results are summarized in Table \ref{tab:results}, which we describe below. It is worth noting that, even though we only report the bias associated to the parameter listed in this table, neglecting lensing magnification and GR effects also leads to biases in other standard $\Lambda$CDM parameters. We do not report these here, and rather concentrate on the parameter spaces that future large-scale structure facilities will target specifically.

\begin{table*}[t]
\begin{tabular}{|l|c|c|c|c||c|c|c|c||c|}
\hline
 & \multicolumn{3}{c}{{\bf all tracers}} & & \multicolumn{3}{c}{{\bf LSST galaxy clustering}} &&\\
\hline 
{\bf Parameters} & improvement & bias from & improvement & bias from & improvement & bias from & improvement & bias from & Max. error \\
 & on $\sigma$ from & magnification & on $\sigma$ from & GR effects & on $\sigma$ from & magnification & on $\sigma$ from & GR effects & on $s(z)$\\ 
 & magnification &  & GR effects &  & magnification &  & GR effects &&  \\
\hline
\textit{wCDM} & & & & & & & &&\\
$\sum m_\nu$ & $<1$\% & 320\% & $<1$\% & 3\% & 2\% & 255\% & $<1$\% & 3\% &9.8\%\\
$w_a$ & $<1$\% & -203\% & $<1$\% & $<1$\% & 8\% & 125\% & $<1$\% & -2\%&5.6\%\\
$w_0$ & $<1$\% & 261\% & $<1$\%  & -3\% & 2\% & -269\% & $<1$\% & 6\%&4.2\%\\
\hline 
\hline
\textit{Horndeski} & & & & & & & &&\\
$c_{\rm M}$ & $<1$\% & 175\% & $<1$\% & -7\% & 3\% & 139\% & $<1$\% & $<1$\%&22\%\\
$c_{\rm B}$ & $<1$\% & 573\% & $<1$\% & 1\% & 76\% & 104\% & $<1$\% & -5\%&11\%\\
$c_{\rm T}$ & $<1$\% & -237\% & $<1$\% & 8\% & 7\% & -66\% & $<1$\% &$<1\%$&23\%\\
\hline
\hline
\textit{non-Gauss.} & & & & & & & &&\\
$f_{\rm NL}$ & -2\% & 17\% & -3\% & -45\% & -2\% & 168\% & -6\% & 7\%&N.A.\\
\hline
\end{tabular} 
\caption{\label{tab:results} Summary of results: improvement on the 1$\sigma$ uncertainties, and parameter bias associated to the contributions of lensing magnification and GR effects to the number counts power spectrum. The left set of columns corresponds to the combination of all tracers (LSST clustering, LSST shear and S4), while the right columns correspond to LSST clustering only. Note that all results are shown as a relative improvement or bias, normalized by the fiducial 1$\sigma$ uncertainties (which are different in these two cases). The three sets of rows correspond to the three parameter families studied here: wCDM+$m_\nu$ (top), Horndeski models (middle) and primordial non-Gaussianity (bottom). Note also that the constraints on $f_{\rm NL}$ and Horndeski models are also marginalized over $\Sigma m_\nu$. The last column shows, for all tracers jointly, the maximum systematic error on $s(z)$ that can be allowed to avoid a bias on each parameter larger than its $1\sigma$ uncertainty.}
\end{table*}

\begin{figure}
  \centering
  \includegraphics[width=0.49\textwidth]{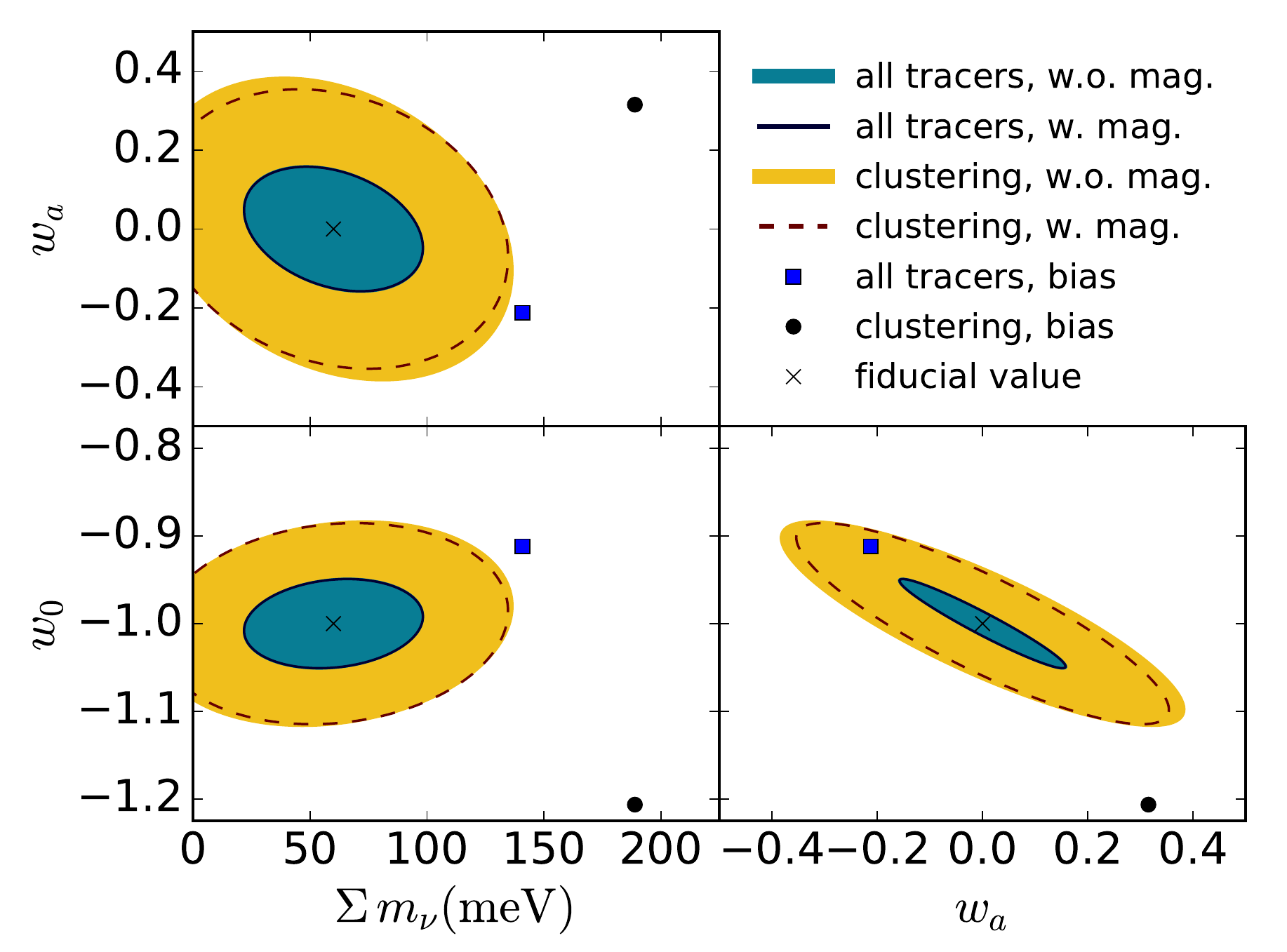}
  \caption{Forecast $1\sigma$ contours for $\Sigma m_\nu$, $w_0$ and $w_a$ from LSST clustering only (orange ellipses) and LSST clustering $+$ LSST shear $+$ S4 (cyan ellipses) in the fiducial case without lensing magnification or GR effects. The thin solid and dashed ellipses correspond to the $1\sigma$ contours after including the lensing contribution to the clustering power spectrum in the same two cases respectively. The black circle and square show the bias associated with ignoring the presence of lensing magnification (again in the same two cases). In all cases the impact of GR effects is negligible, and therefore we have not included the corresponding ellipses in this figure.}\label{fig:res.wmnu}
\end{figure}

\subsection{Impact on dark energy and neutrino mass}\label{ssec:results.wmnu}
  As has been previously shown by \cite{Cardona:2016qxn}, neglecting the lensing magnification effect can significantly bias the estimation of the total sum of neutrino masses $\sum m_\nu$. Our analysis here extends this study to the dark energy equation of state parameters, $w_0$ and $w_a$, since they have been shown to be degenerate with $\sum m_\nu$~\cite{Allison:2015qca} (see also \cite{2014MNRAS.437.2471D}, where $w_0$ and $w_a$ were studied independently of $\sum m_\nu$).
  
  In addition to this, the combined analysis of galaxy clustering and cosmic shear data is known to be of great use in breaking degeneracies to constrain these parameters \cite{desresults}. This is relevant for two reasons: on the one hand, it is worth exploring to what extent the lensing information contained within the magnification bias contribution to galaxy clustering can also be used to break these same degeneracies {\it in lieu} of cosmic shear \cite{2014MNRAS.437.2471D}. On the other hand, since cosmic shear is a direct probe of gravitational lensing, it is interesting to study whether any biases associated with neglecting the magnification bias term could be mitigated by including cosmic shear information.
  
  Finally, although the large-scale relativistic effects are known to be barely measurable, our treatment will allow us to explore their impact on constraints and systematic biases.
  
  The results are shown in Figure \ref{fig:res.wmnu}, where the orange ellipses show the $1\sigma$ contours using only clustering information from LSST and the cyan ellipses correspond to the full constraining power of LSST clustering, LSST shear and S4 (including primary CMB and lensing). The thin solid and dashed ellipses correspond to the constraints after accounting for the contribution of magnification to clustering in the same two cases respectively. Although using only clustering information the magnification term does improve constraints slightly (up to $8\%$ in the marginalized uncertainties), the improvement is absolutely negligible when
  including all other cosmological probes.
  
  In the same plots, the filled circles and squares show the forecast bias on the same parameters, both for clustering alone and including all probes respectively. Although the inclusion of CMB and shear data reduces the size of the bias, the faster improvement in the constraints makes the significance of this bias worse. It is worth pointing out that the direction of the bias changes after including new probes, due to the change in direction of the different degeneracies.
  
  We have also evaluated the information content (i.e. improvement in constraints) of the GR terms as well as the parameter bias they induce. The information content is completely negligible, with an improvement in the 1$\sigma$ uncertainties well below $1\%$ in all cases. The bias associated with the omission of these terms is equally negligible, with a maximum fractional bias of $6\%$ with respect to the standard deviation in the case of $w_0$ when only galaxy clustering data are taken into account. These biases are further suppressed when including other probes.
  
  \begin{figure}
    \centering
    \includegraphics[width=0.49\textwidth]{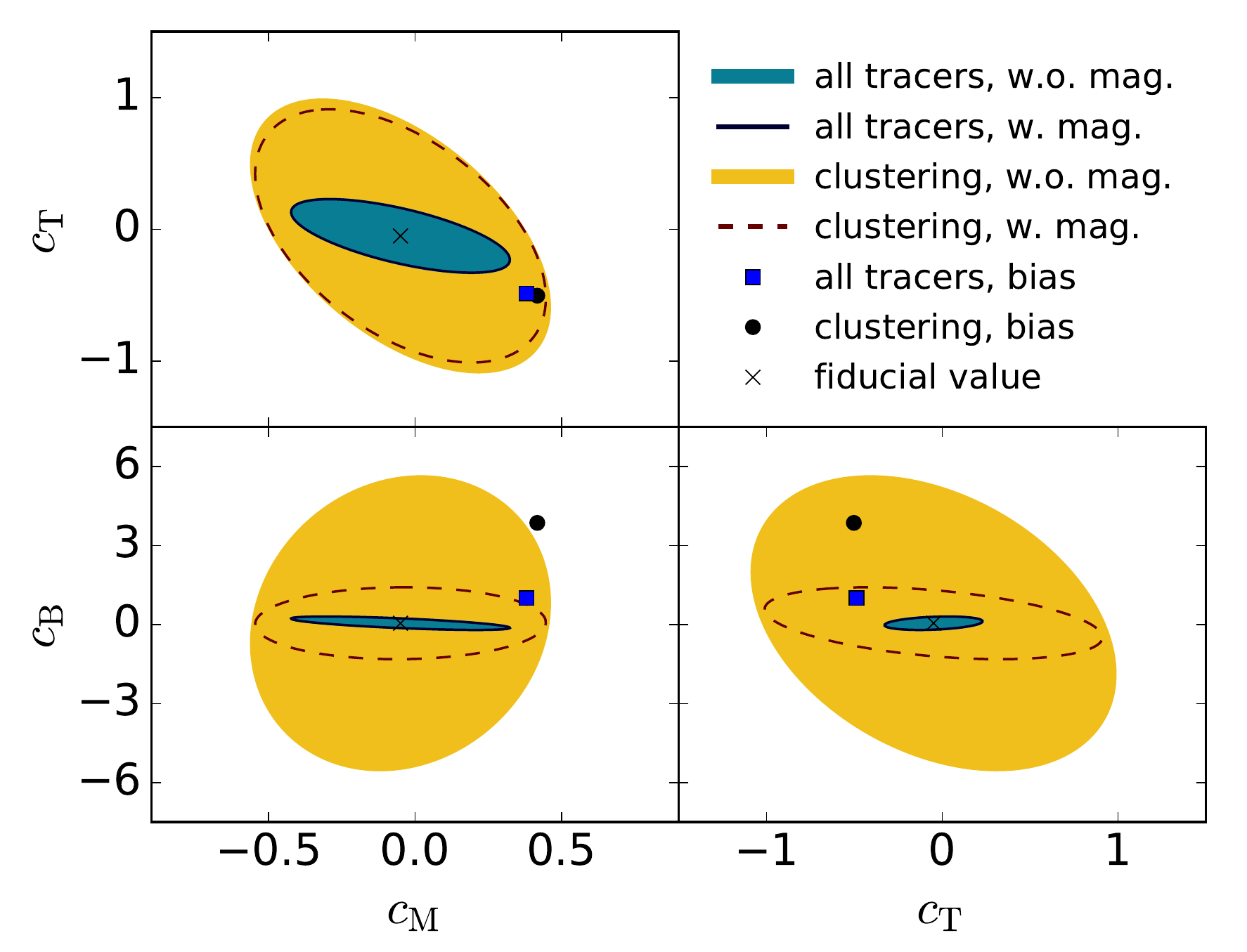}
    \caption{Same as Figure \ref{fig:res.wmnu} for the Horndeski parameters $c_B$, $c_M$ and $c_T$.}\label{fig:res.horndeski}
  \end{figure}
\subsection{Impact on scalar-tensor theories}\label{ssec:results.horndeski}
  As shown in the previous section, the secondary clustering anisotropies (lensing and GR effects) do not contain significant extra information in terms of final constraints on cosmological pararameters for standard departures from vanilla $\Lambda$CDM. One could however argue that the true constraining power of these relativistic terms would be realized on actual modifications of GR \cite{2016JCAP...07..040R}, and therefore it is relevant to explore this possibility. To that end we have repeated the same Fisher analysis on the Horndeski parametrization of scalar-tensor gravity theories described in Section \ref{ssec:methodology.parameters}.
  
  The results are shown in Figure \ref{fig:res.horndeski} using the same color coding as Fig. \ref{fig:res.wmnu}. Interestingly, when including only clustering information we observe a large improvement in the constraint on $c_B$, and no real improvement on $c_T$ and $c_M$. An inspection of the correlation coefficients between different parameters reveals that the inclusion of magnification is able to break strong degeneracies between $c_B$ and the nuisance galaxy bias parameters, as could have been expected given that lensing effects trace the dark matter perturbations directly, and therefore marginally help constraint $b(z)$. In all cases, the bias associated with the lensing term is of the same order as the $1\sigma$ uncertainty when using only clustering information, smaller than the case explored in the previous section. These results change, however, when all probes are included simultaneously: the relative constraining power of the magnification term becomes negligible in the presence of cosmic shear and CMB, while the improvement in the
  final constraints brought about by these probes makes the bias associated to the lensing term significant at the $5\sigma$ level for $c_B$.
  
  Regarding the relevance of the other GR effects, we find the same results obtained in the previous sections: these terms do not significantly improve the final constraints on the Horndeski parameters ($<1\%$), and do not induce a significant bias ($\sim8\%$ of $\sigma$ at worst).

  \begin{figure}
    \centering
    \includegraphics[width=0.49\textwidth]{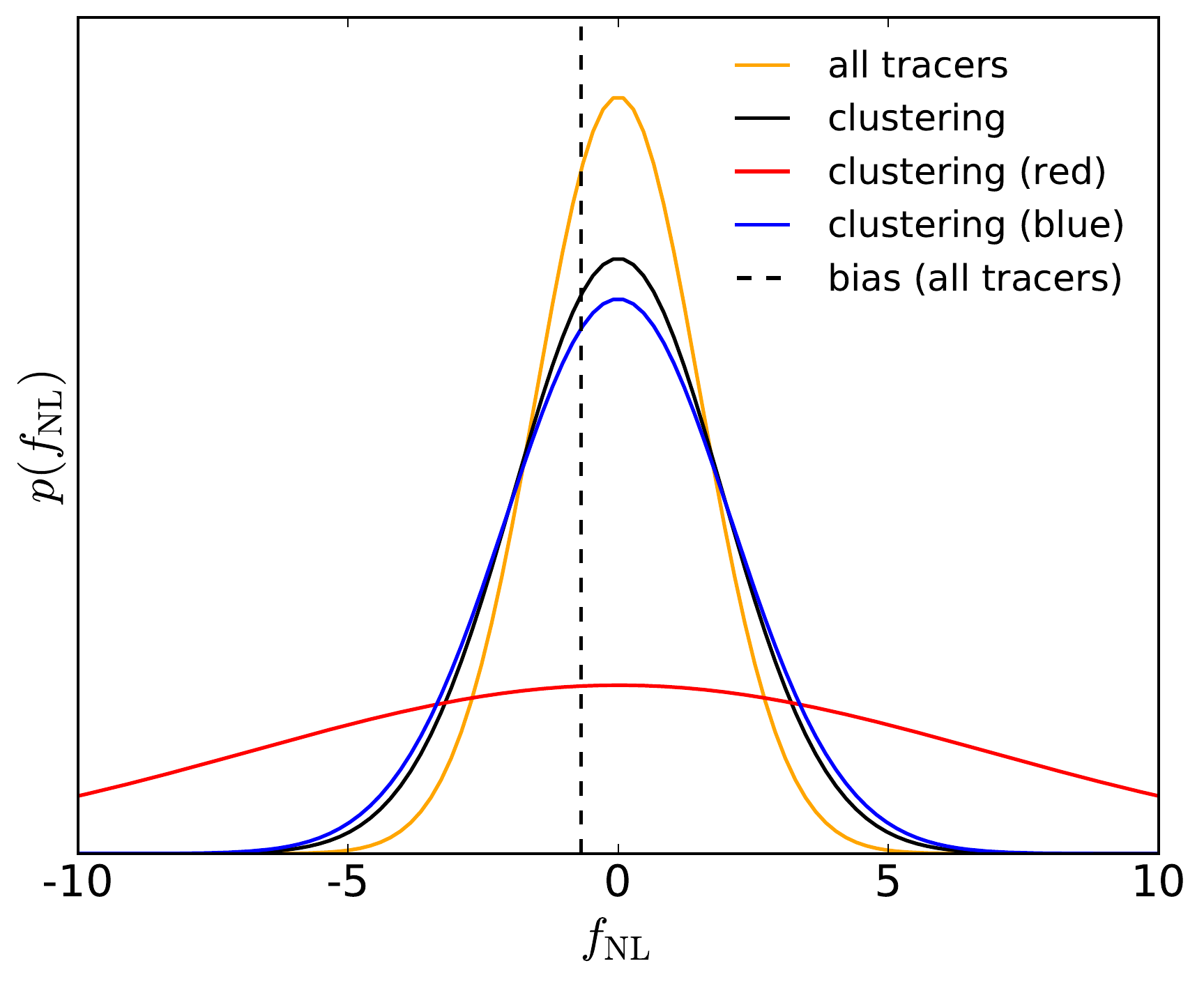}
    \caption{Forecast distribution for $f_{\rm NL}$ for LSST red galaxies (red), blue galaxies (blue), the combination of both in a multi-tracer sense (black) and the combination of LSST galaxy clustering, LSST cosmic shear and CMB S4 (orange). The bias on $f_{\rm NL}$ associated with the GR effects, corresponding to $f^{\rm GR}_{\rm NL}\simeq-0.7$ is shown as a vertical dashed line}\label{fig:res.fnl}
  \end{figure}
\subsection{Impact on primordial non-Gaussianity}\label{ssec:results.fnl}
  Except for the magnification lensing term, all other relativistic corrections to the number counts power spectrum dominate on horizon-sized scales. Therefore, although these effects seem to be irrelevant on the standard cosmological parameters explored in the previous sections, any parameter sensitive to the clustering pattern on large scales may be more affected by them. This is the case for the effects of primordial non-Gaussianity on the clustering pattern of biased tracers, as discussed in Section \ref{ssec:methodology.parameters}. We have therefore carried out the same Fisher analysis done in Section \ref{ssec:results.wmnu} including $f_{\rm NL}$ as a free parameter.
  
  The results are shown in Figure \ref{fig:res.fnl} as 1D posterior distributions for $f_{\rm NL}$ marginalized over all other parameters (including $w_0$, $w_a$ and $\Sigma m_\nu$). Before discussing the relevance of the lensing and GR effects it is worth inspecting the improvement on $\sigma(f_{\rm NL})$ from the inclusion of different probes. Here we have considered the cases of the blue and red clustering samples individually, the combination of both and the addition of external datasets (weak lensing and CMB data). For the blue and red samples, as well as their combination, we recover the same result obtained in \cite{Alonso:2015sfa}: the red sample alone does not yield competitive constraints given its small volume coverage ($\sigma(f_{\rm NL}|{\rm red})\simeq7$), while the higher number density and volume of the blue galaxies allows for a more interesting bound ($\sigma(f_{\rm NL}|{\rm blue})\simeq2$). The combination of both samples yields a slightly better constraint due to the multi-tracer effect, and the addition of external datasets improves it further $\sigma(f_{\rm NL}|{\rm all\,tracers})\simeq1.5$, mostly due to the improved measurement of the galaxy bias.
  
  When switching on the lensing and GR effects we observe no significant improvement or degradation in $\sigma(f_{\rm NL})$. On the other hand we observe that GR effects cause a bias of $\sim50\%$ for the combination of all tracers\footnote{Note that this value is found after marginalizing over all other cosmological and nuisance parameters. However, the result holds also under the assumption that all parameters other than $f_{\rm NL}$ are known.}, corresponding to an effective value of $f_{\rm NL}^{\rm GR}\simeq-0.7$. This is in agreement with \cite{2012JCAP...10..025B,2012PhRvD..85b3504J,2012PhRvD..85d1301B,2017JCAP...03..006D}. Although this may not be a concern for the experimental setup considered here, other experiments targetting $f_{\rm NL}$ explicitly, such as SPHEREx \cite{2016arXiv160607039D}, may need to account for these relativistic corrections. Magnification lensing, on the other hand, causes a much smaller effect, given its scale dependence. In the absence of CMB or cosmic shear measurements, we observe however a large bias on $f_{\rm NL}$ (of order 1$\sigma$) induced by magnification lensing. This is caused by the biased estimation of the galaxy bias parameters, which affect the amplitude of the correction due to $f_{\rm NL}$ if magnification is not taken into account (see also \cite{2011PhRvD..83l3514N}).
  
\subsection{Impact of magnification uncertainties}
  In the previous sections we have seen that the magnification term is important and can significantly bias cosmological parameter estimates if unaccounted for, as has also been previously shown by Refs.~\cite{2014MNRAS.437.2471D,Cardona:2016qxn}. Since the amplitude of the magnification term depends on the slope of the source number counts with apparent magnitude (see Eq.~\ref{eq:sz}), an outstanding question is how well $s(z)$ needs to be measured in order to avoid a significant bias ($>1\sigma$) from the magnification-related uncertainties alone. In order to test this, we have recomputed our forecasts for both the $\mathrm{wCDM}+\Sigma\, m_\nu$ and Horndeski models, this time using a theoretical power spectrum that includes magnification bias with our fiducial model for $s(z)$, and an observed power spectrum in which we increase $s(z)$ by $10\%$. This then allows us to estimate the parameter bias associated with a 10\% systematic uncertainty on $s(z)$ using the formalism described in Sec.~\ref{sec:methodology}. We find that the parameters of key relevance for galaxy clustering ($\sum m_\nu,\,w_0,\,w_a,\,c_{\rm B},\,c_{\rm M}$ and $c_{\rm T}$) can be significantly biased by uncertainties of this order (e.g. 179\% of $\sigma$ for $w_0$).
  
  These results can be used to quantify the level to which $s(z)$ must be known to avoid biasing individual parameters. Under the assumption that the parameter bias $\Delta\theta$ scales linearly with the relative systematic error on $s$, $\delta s$, we can estimate $\Delta\theta$ for any $\delta s$ in terms of the bias computed in the 10\% case $\Delta\theta=\rho\,\delta s$, where $\rho\equiv\Delta\theta/\delta s$ for $\delta s=0.1$. Then, assuming that we can at most afford a bias $\Delta\theta=\epsilon\,\sigma(\theta)$, where $\sigma(\theta)$ is the 68\% uncertainty on $\theta$ and $\epsilon\sim O(1)$, the corresponding maximum relative systematic error of $s$ is given by
  \begin{equation}
    \left.\delta s\right|_{\rm max}=\frac{\epsilon}{\rho}\sigma(\theta).
  \end{equation}
  For $\epsilon=1$, the allowed relative uncertainties for the different parameters are given in the last column of Table \ref{tab:results}. We find that $s(z)$ must be correctly determined to the $\sim5\%$ level in order to avoid significant biases on the dark energy parameters and the sum of neutrino masses. For the case of Horndeski parameters this requirement is relaxed to a $\sim10\%$ systematic uncertainty, but we note that, given the degeneracy between the $c_X$ and other standard cosmological parameters such as $h$, a systematic error on $s(z)$ could propagate into these as well. Consistency studies between different sets of probes will therefore be vital to detect these and other types of systematics.

\section{Discussion}\label{sec:discussion}
  \begin{figure}
    \centering
    \includegraphics[width=0.49\textwidth]{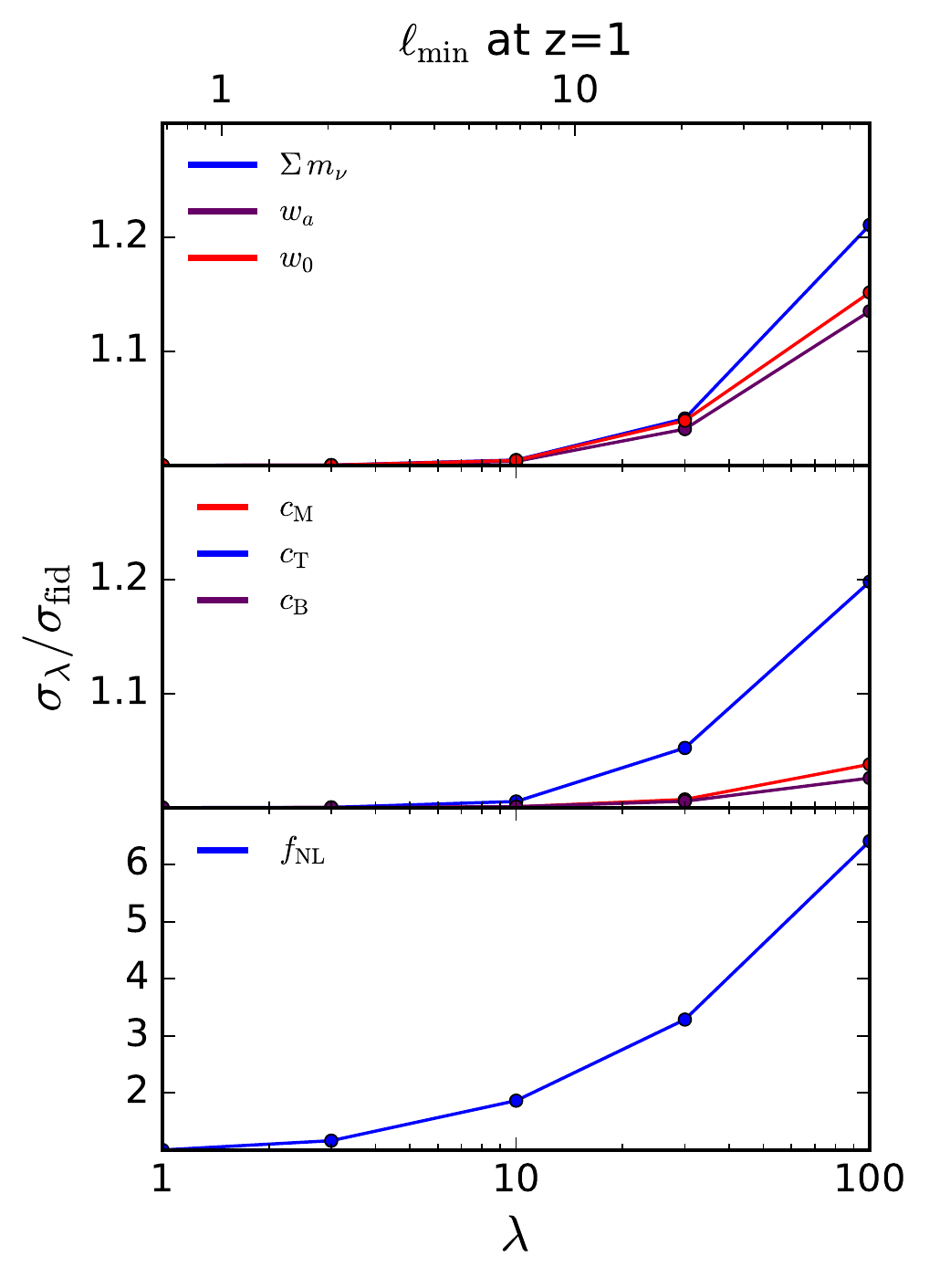}
    \caption{Relative degradation in the final constraints associated with removing all scales larger than a factor $\lambda$ times the comoving horizon at the source redshift (the associated angular scales at $z=1$ are shown in the upper twin $x$-axis). The results are shown for simple extensions to $\Lambda$CDM (upper panel), scalar-tensor theories (middle panel) and primordial non-Gaussianity (lower panel). Except in the case of $f_{\rm NL}$, the information content of the largest scales is heavily suppressed due to cosmic variance.}\label{fig:lambda}
  \end{figure}
  Accurate measurements of the large scale structure of the Universe are the  next frontier of modern cosmology. Maps of the galaxy and diffuse gas distributions, of the CMB and of the gravitational potential via weak lensing will be used to place tight constraints on a plethora of cosmological parameters. In the past few years we have learnt of the importance of taking into account novel corrections to the observables of large scale structure, specifically through lensing magnification and GR effects. In this paper, we have investigated how important these secondary corrections to the power spectrum of galaxy number counts are in terms of information content and potential biases to cosmological parameters. We have explored the relevance of these effects on three different families of cosmological parameters: extensions to the standard $\Lambda$CDM paradigm in the form of massive neutrinos and time-varying dark energy equation of state, Horndeski-like parametrizations of scalar-tensor theories, and the large-scale contribution of primordial non-Gaussianity to the galaxy power spectrum.

  It is natural to split the secondary contributions mentioned above into two classes: the contribution from lensing magnification is relevant on small angular scales and is coherent over large redshift separations. This contribution is well known and has been used in the past in different scientific analyses. We group all other contributions under the umbrella term of ``GR effects'', given their relevance mostly on large scales, of the order of the horizon at the redshift of the source.

  We have established, in agreement with previous studies \cite{2014MNRAS.437.2471D,Cardona:2016qxn}, that even though lensing magnification can be detected with high significance, it will not in general contribute strongly to improve the final constraints on any cosmological parameter. Although it may be relevant to constrain deviations from modified gravity (e.g. $c_B$ in Section \ref{ssec:results.horndeski}) using only clustering data, its information content is negligible when combined with cosmic shear and CMB observations. Nevertheless, using a Fisher approach we have shown that it will be necessary to model and account for this contribution to the galaxy power spectrum in order to avoid strong biases on dark-energy parameters and the sum of neutrino masses. The bias associated with neglecting the effects of magnification is most relevant when considering clustering alone as a cosmological probe, and gets reduced considerably after including shear and CMB. The reduced parameter uncertainties in the latter case imply that the associated biases are still significant, however. Our approach also allows us to quantify the level to which the number counts slope $s(z)$ must be known in order to avoid significantly biasing the most relevant late-time cosmological parameters. We find that $s(z)$ must be known at least the $\sim5\%$ level, in rough agreement with \cite{2014MNRAS.437.2471D}. An MCMC-based approach will be able to fully test the extent of these biases in a realistic scenario. On the other hand, and as expected given the scale dependence of the lensing contribution, this effect should not have a strong impact on the inferred value of $f_{\rm NL}$ given expected uncertainties.

  The GR effects, on the other hand, are known to have a sub-dominant amplitude and, as expected, we find that they will have a negligible impact on both the uncertainty and bias on most cosmological parameters. The only exception to this is the level of primordial non-Gaussianity, given the similar scale dependence of these effects and the $\sim1/k^2$ contribution of $f_{\rm NL}$. We find that the GR effects could induce a bias on this parameter of the order of $f_{\rm NL}^{\rm GR}\sim0.7$, in agreement with previous estimate of the amplitude of these contributions. This is comparable to the uncertainty on $f_{\rm NL}$ expected from LSST, and will therefore be relevant for future experiments specifically targeting this science case. We emphasize though that systematic effects that may cause correlated fluctuations in the homogeneity of the galaxy sample (e.g. depth variations, dust extinction, star contamination) will need to be carefully treated in order to minimize their impact on the large-scale galaxy power spectrum, thus preserving this sensitivity of galaxy surveys to $f_{\rm NL}$. Since the amplitude of the GR effects depends on the value of the magnification and evolution biases $s(z)$ and $f_{\rm evo}(z)$, the uncertainties on these quantities may hamper our ability to reach optimal constraints on $f_{\rm NL}$ or mitigate the associated bias on this parameter. This underpins the need to quantify the luminosity and time dependence of the background number density of sources for future Stage-IV surveys, already noted in the literature (e.g. \cite{2014MNRAS.437.2471D}) in the context of the impact of lensing magnification on standard cosmological parameters.

  It is also worth mentioning that, even though these GR effects are one of the few manifestly relativistic contributions to the power spectrum, and therefore may potentially contain valuable information to constrain departures from General Relativity, we find that their constraining power on modified gravity theories is negligible. This can be easily understood in terms of the scales involved: even if a given modified gravity theory could generate a significant difference in any of these GR terms, these effects are only relevant on horizon-size scales, and therefore their information content is heavily suppressed by cosmic variance. This can be explicitly verified by re-running these forecasts cutting out the largest scales and comparing the results with our fiducial predictions. To do so, for each redshift bin $i$ with a median redshift $z_i$, we define a minimum scale $\ell_{\rm min}(z_i,\lambda)$ as the Fourier scale corresponding to the angular size of the horizon at that redshift divided by a factor $\lambda$:
  \begin{equation}
    \ell_{\rm min}(z_i,\lambda)=\lambda\,\chi(z_i)\frac{H(z_i)}{1+z_i}
  \end{equation}
  Figure \ref{fig:lambda} shows the increment in the uncertainty of different cosmological parameters associated with the loss of these large scales as a function of $\lambda$. The results were obtained for the combination of LSST clustering, shear and CMB S4. We observe that, even removing scales that are 1\% the size of the horizon, the degradation in the final constraints is at most $\sim20\%$ for all cosmological parameters, with the exception of $f_{\rm NL}$. 

  On a different front, one might hope that the inclusion of the lensing magnification term in the number counts might mitigate some systematic uncertainties - specifically, it might help to pin down galaxy bias. And indeed,  in the analysis of Horndeski theories, we have shown that including that term significantly changes the uncertainty in $c_B$ by breaking some of its degeneracies with the galaxy bias parameters. While this is the case, it is not accompanied by a substantial reduction in the uncertainties in these parameters; the reduction in the uncertainty is of the order of a few percent.

  One interesting aspect that we have not explored is the importance of the effects studied here in cross-correlations between the Integrated Sachs-Wolfe (ISW) effect and number counts \cite{1996PhRvL..76..575C}. This measurement could be particularly relevant to constrain modified gravity theories \cite{2017arXiv170702263R}. In principle the non-inclusion of the GR terms could bias estimates of cosmological parameters although this should strongly depend on the scales which are included in the standard analysis. We leave a systematic analysis of the ISW effect for future work.

  Finally, it is worth stressing the fact that the results presented here are applicable to the combination of CMB and photometric galaxy samples assumed. Spectroscopic surveys, on the other hand, might be able to detect some of the GR effects studied here on intermediate scales as a local dipole in the cross-correlation function of different galaxy samples \cite{2014PhRvD..89h3535B}. Although this is a challenging measurement \cite{2017JCAP...01..032G,2017MNRAS.470.2822A}, it would be important to further understand its constraining power.
  
  With this paper we have assessed the importance of relativistic effects on cosmological parameter estimation with a particular emphasis on extensions of $\Lambda$CDM. Our understanding of the impact of these effects on the analysis of future data will allow us to reap the rewards of the  next generation of cosmological surveys.

\section*{Acknowledgements}
The authors would like to thank Miguel Zumalac\'arregui, Roy Maartens, Christopher Duncan, Erminia Calabrese and Emilio Bellini for useful comments and discussions. We also thank the authors of \cite{Cardona:2016qxn} (and W. Cardona and F. Montanari in particular) for constructive exchanges about their analysis. CSL is supported by a Clarendon Scholarship and acknowledges support from Pembroke College, Oxford. DA is supported by the STFC and the Beecroft Trust. PGF acknowledges support from Leverhulme, STFC, the Beecroft Trust and the ERC.



\bibliographystyle{apsrev4-1}
\bibliography{paper}

\appendix
\onecolumngrid
\section{Complete expressions for the corrections to the number counts of galaxies}
  \label{app:GRterms}
  The linear-order expression for the transfer function of number-count fluctuations in the $i$-th redshift bin, characterized by a radial selection function $W_i(z)$, is given as a sum over 10 different terms \cite{DiDio:2013bqa}:
  \begin{align}\label{eq:terms0}
    &\Delta^{{\rm D},i}_\ell(k)\equiv\int d\eta \,b\,\tilde{W}_i\,
    \delta_{M}(k,\eta)\,j_\ell(k\chi(\eta)),\hspace{12pt}
    ~~~~~~~~\Delta^{{\rm RSD},i}_\ell(k)\equiv\int d\eta\,(aH)^{-1}
    \tilde{W}_i(\eta)\,\theta(k,\eta)\,
    j_\ell''(k\chi(\eta)),\\
    &\Delta^{{\rm L},i}_\ell(k)\equiv \ell(\ell+1) \int d\eta\,
    \tilde{W}^{\rm L}_i(\eta)\,
    (\phi+\psi)(k,\eta)\,j_\ell(k\chi(\eta)),\hspace{12pt}
    \Delta^{{\rm V1},i}_\ell(k)\equiv \int d\eta\,(\fevo-3)\,aH\,\tilde{W}_i(\eta)\,
    \frac{\theta(k,\eta)}{k^2}\,j_\ell(k\chi(\eta)),\\
    &\Delta^{{\rm V2},i}_\ell(k)\equiv \int d\eta\,
    \left(1+\frac{H'}{aH^2}+\frac{2-5s}{\chi\,aH}+5s-\fevo\right)
    \tilde{W}_i(\eta)\,\frac{\theta(k,\eta)}{k}\,j_\ell'(k\chi(\eta)),\\
    &\Delta^{{\rm P1},i}_\ell(k)\equiv \int d\eta\,
    \left(2+\frac{H'}{aH^2}+\frac{2-5s}{\chi\,aH}+5s-\fevo\right)
    \tilde{W}_i(\eta)\,\psi(k,\eta)\,j_\ell(k\chi(\eta)),\\
    &\Delta^{{\rm P2},i}_\ell(k)\equiv \int d\eta\,
    (5s-2)\tilde{W}_i(\eta)\,\phi(k,\eta)\,j_\ell(k\chi(\eta)),\hspace{12pt}
    \Delta^{{\rm P3},i}_\ell(k)\equiv \int d\eta\,
    (aH)^{-1}\tilde{W}_i(\eta)\,\phi'(k,\eta)\,j_\ell(k\chi(\eta)),\\\label{eq:isw}
    &\Delta^{{\rm P4},i}_\ell(k)\equiv \int d\eta\,
    \tilde{W}^{\rm P4}_i(\eta)\,(\phi+\psi)(k,\eta)\,j_\ell(k\chi(\eta)),\hspace{12pt}
    \Delta^{{\rm ISW},i}_\ell(k)\equiv \int d\eta\,
    \tilde{W}^{\rm ISW}_i(\eta)\,(\phi+\psi)'(k,\eta)\,j_\ell(k\chi(\eta)).
  \end{align}
  Here, $j_\ell(x)$ is the spherical Bessel function of order $\ell$, and we have defined the window functions
  \begin{align*}\label{eq:windows}
    &\tilde{W}_i(\eta(z))\equiv W_i(z)\left(\frac{d\eta}{dz}\right)^{-1},\hspace{12pt}
    \tilde{W}^{\rm L}_i(\eta)\equiv\int_0^\eta d\eta'\tilde{W}_i(\eta')
    \frac{2-5s(\eta')}{2}\frac{\chi(\eta)-\chi(\eta')}{\chi(\eta)\chi(\eta')},\\
    &\tilde{W}^{\rm P4}_i(\eta)\equiv\int_0^\eta d\eta'\tilde{W}_i(\eta')
    \frac{2-5s}{\chi},\hspace{12pt}
    \tilde{W}^{\rm ISW}_i(\eta)\equiv\int_0^\eta d\eta'\tilde{W}_i(\eta')
    \left(1+\frac{H'}{aH^2}+\frac{2-5s}{\chi\,aH}+5s-\fevo\right)_{\eta'}.
  \end{align*}
  The quantities $\delta_{M}$, $\theta$, $\phi$ and $\psi$ above are transfer functions for density perturbations in the comoving synchronous gauge, for the velocity divergence in the conformal Newtonian gauge and for two metric potentials in the same gauge\footnote{The conformal Newtonian gauge is defined by the line element $ds^2=-a^2(\eta)\,[(1+2\psi)d\eta^2-(1-2\phi)\delta_{ij}dx^idx^j\,]$.}.
  
  Of the 10 terms in Eq. \ref{eq:terms0} above, $\Delta^{\rm D}$, $\Delta^{\rm RSD}$ are the dominant density and redshift-space distortions terms respectively, $\Delta^{\rm L}$ is the contribution of lensing magnification and we have grouped the remaining 7 terms under 
  a single ``GR effects'' contribution $\Delta^{\rm GR}$ in Eq. \ref{eq:terms1}.
  
\section{Details of the bias calculation}
  \label{app:Fisher bias calculation}
  In section \ref{ssec:methodology.fisher}, we approximate $\langle\partial_\alpha\partial_\beta\chi^2(\theta_{\rm obs})\rangle$ as ${\sf F}_{\alpha\beta}$ around the inferred (and possibly biased) parameters (here $\chi^2\equiv-2\log{\cal L}$). We show here why this approximation is valid at the linear level. Differentiating equation \ref{eq:like}, we find 
  \begin{align}
    \partial_\alpha\chi^2=\sum_\ell f_\mathrm{sky}\frac{2\ell+1}{2}&\left[{\rm Tr}({\sf C}_\ell^{-1}\partial_\alpha{\sf C}_\ell)-\sum_m\frac{{\bf a}_{\ell m}^\dag\,{\sf C}_\ell^{-1}\partial_\alpha{\sf C}_\ell{\sf C}_l^{-1}{\bf a}_{\ell m}}{2\ell+1}\right]\\\nonumber
    \partial_\alpha\partial_\beta\chi^2=\sum_\ell f_\mathrm{sky}\frac{2\ell+1}{2}&\left[{\rm Tr}({\sf C}_\ell^{-1}\partial_\alpha\partial_\beta{\sf C}_\ell)-{\rm Tr}(\partial_\alpha{\sf C}_\ell{\sf C}_\ell^{-1}\partial_\beta{\sf C}_\ell{\sf C}_\ell^{-1})-\sum_m\frac{{\bf a}_{\ell m}^\dag\,{\sf C}_\ell^{-1}\partial_\alpha\partial_\beta{\sf C}_\ell{\sf C}_l^{-1}{\bf a}_{\ell m}}{2\ell+1}\right.\\
    &\left.\,\,+\sum_m\frac{{\bf a}_{\ell m}^\dag\,{\sf C}_\ell^{-1}\partial_\alpha{\sf C}_\ell{\sf C}_l^{-1}\partial_\beta{\sf C}_\ell{\sf C}_l^{-1}{\bf a}_{\ell m}}{2\ell+1}+\sum_m\frac{{\bf a}_{\ell m}^\dag\,{\sf C}_\ell^{-1}\partial_\beta{\sf C}_\ell{\sf C}_l^{-1}\partial_\alpha{\sf C}_\ell{\sf C}_l^{-1}{\bf a}_{\ell m}}{2\ell+1}\right]
  \end{align}
  Since $\langle{\bf a}^\dag\,{\bf a}\rangle={\sf C}_\ell^{\rm obs}$, we find the expectation value
  \begin{equation}
    \langle\partial_\alpha\chi^2\rangle=-\sum_\ell f_\mathrm{sky}\frac{2\ell+1}{2}{\rm Tr}({\sf C}_\ell^{-1}\partial_\alpha{\sf C}_\ell{\sf C}_\ell^{-1}\Delta{\sf C}_\ell),
  \end{equation}
  where we have defined ${\sf C}_\ell^{\rm obs}\equiv{\sf C}_\ell+\Delta C_\ell$. This yields the expression for the vector ${\bf v}$ given in equation \ref{eq:v}. For the second derivatives we find
  \begin{align}\label{eq:fishb}
    \langle\partial_\alpha\partial_\beta\chi^2(\theta_{\rm obs})\rangle=\sum_\ell f_\mathrm{sky}\frac{2\ell+1}{2}\Big[{\rm Tr}({\sf C}_\ell^{-1}\partial_\alpha{\sf C}_\ell{\sf C}_\ell^{-1}\partial_\beta{\sf C}_\ell)+{\rm Tr}({\sf K}_{\alpha\beta,\ell}\,\Delta{\sf C}_\ell)\Big]
  \end{align}
  where ${\sf K}_{\alpha\beta,\ell}$ is given by
  \begin{align}
    {\sf K}_{\alpha\beta,\ell}\equiv{\sf C}_\ell^{-1}\partial_\alpha{\sf C}_\ell{\sf C}_\ell^{-1}\partial_\beta{\sf C}_\ell{\sf C}_\ell^{-1}+{\sf C}_\ell^{-1}\partial_\beta{\sf C}_\ell{\sf C}_\ell^{-1}\partial_\alpha{\sf C}_\ell{\sf C}_\ell^{-1}-{\sf C}_\ell^{-1}\partial_\alpha\partial_\beta{\sf C}_\ell{\sf C}_\ell^{-1}.
  \end{align}
  Therefore, if $\Delta{\sf C}_\ell\approx\partial_\alpha{\sf C}_l\cdot(\theta^{\rm obs}_\alpha-\theta^{\rm th}_\alpha)$, the second term in Eq. \ref{eq:fishb} is of second order and we can approximate $\langle\partial_\alpha\partial_\beta\chi^2(\theta_{\rm obs})\rangle$ as ${\sf F}_{\alpha\beta}$. 

\twocolumngrid

\end{document}